\documentclass[twocolumn,aps,pra,a4paper,showpacs,preprintnumbers,amsmath,amssymb]{revtex4}

\usepackage[english]{babel}
\usepackage[dvips]{graphicx}% Include figure files
\usepackage{dcolumn}% Align table columns on decimal point
\usepackage{bm}% bold math
\usepackage{amssymb}
\usepackage{mathrsfs}
\begin{document}
\title{Quantum state tomography of molecular rotation}

\author{Anders S. Mouritzen\footnote{Corresponding author}}
 \email{asm@phys.au.dk}
% \affiliation{University of Aarhus, Department of Physics and Astronomy, University of
%Aarhus, DK-8000 \AA rhus C, Denmark}
\author{Klaus M\o lmer}
 \email{moelmer@phys.au.dk}
\affiliation{QUANTOP, Danish National Research Foundation Center for
Quantum Optics, Department of Physics and Astronomy, University of
Aarhus, DK-8000 \AA rhus C, Denmark}

\date{\today}

\begin{abstract}
We show how the rotational quantum state of a linear or symmetric
top rotor can be reconstructed from finite time observations of the
polar angular distribution under certain conditions. The presented
tomographic method can reconstruct the complete rotational quantum
state in many non-adiabatic alignment experiments. Our analysis
applies for measurement data available with existing measurement
techniques.
\end{abstract}


 %Use showkeys class option if keyword
                              %display desired
\maketitle

\section{\label{sec:intro}Introduction}
Recently, there has been great interest in non-adiabatic alignment
of molecules using short non-resonant laser pulses \cite{vrakking},
for a recent review see \cite{henriktamarreview}. Here the molecules
are excited in a rotational state showing time-dependent angular
anisotropy with ensuing revival structure long after the passage of
the aligning pulse. The time-dependent angular distribution after
passage of the laser pulse can be measured \cite{rotrevival},
\cite{enafhenrik}, and the states created have many uses in
ultra-fast optics, high harmonic generation, scattering theory and
potentially in chemical investigations \cite{henriktamarreview},
\cite{corkumnature}.

So far, the comparison of such experiments with theory has been
though the resemblance of experimental data with numerical
simulations of the time evolution from a known initial state.
Conversely, we propose a tomographic method by which one can find
the initial quantum rotational state of a linear rotor from
measurements of the angular distributions at several points of time.

\begin{figure}[htb!]
\includegraphics[width=0.5\textwidth]{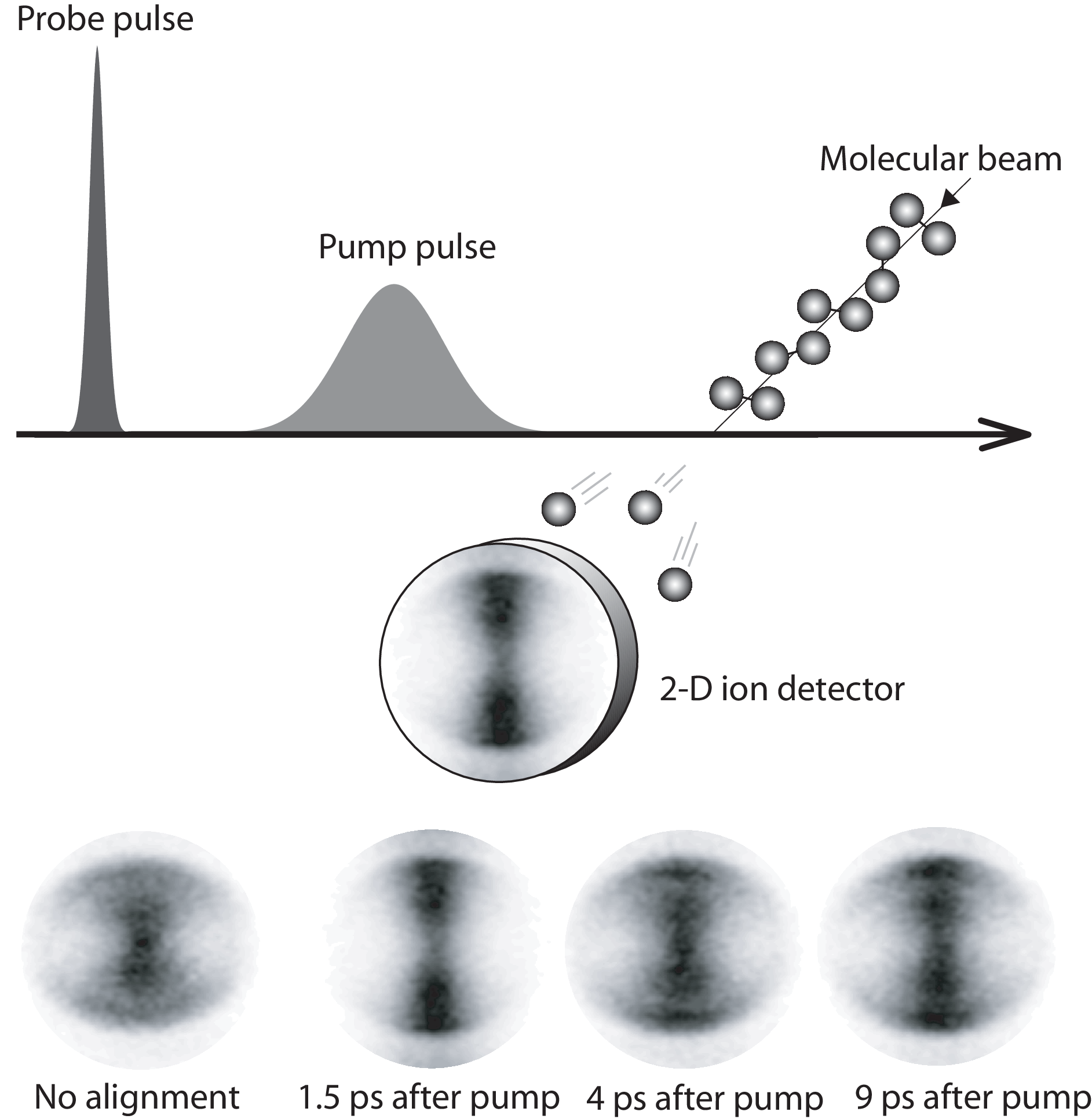}
\caption{The figure shows a typical alignment experiment with short
laser pulses. The cold thermal molecules are excited in a rotational
state by a non-resonant pump pulse vertically polarized. After a
time $t$ the molecules are coulomb-exploded with an intense,
ultrashort probe pulse, also vertically polarized. The fragments
recoil back-to-back and are imaged on a 2-D detector. The 3-D
angular distributions can be found from these 2-D images because of
the cylindrical symmetry. Four examples of experimental data with
Iodobenzene are shown, where the imaged fragments are iodine ions:
(from left) Probe applied long before pump, probe applied after
respectively $1.5$, $4$ and $9$ps. Part of the anisotropy, which is
directly apparent in the un-aligned case, is due to the
coulomb-explosion preferentially occurring parallel to the probe
polarization. This is corrected for when finding the 3-D
distribution. Strictly speaking, Iodobenzene is an asymmetric top,
but only slightly. For many purposes, it behaves like a symmetric
top. Experimental data kindly provided by Simon S. Viftrup.
\label{fig:rotfig}}
\end{figure}

A typical example of an alignment experiment is shown in figure
\ref{fig:rotfig}. Here, a rotationally cold supersonic molecular
beam is used as the molecular system in a pump-probe experiment. The
initial state is thus approximately a thermal state with only a few
energy levels excited. The pump is a short, with a typical duration
of a few picoseconds, intense non-resonant laser pulse, which
induces the rotational state. The pump pulse is linearly polarized
and since the initial state is thermal and therefore rotationally
symmetric, the rotational state formed will be cylindrically
symmetric around the pump polarization axis. Furthermore, the field
mainly interacts with the molecule through the molecular
polarizability, making the rotational state reflection symmetric in
a plane orthogonal to the pump polarization. This last symmetry is
not, however, a requirement for the tomographic method below.

After a delay, the molecules are coulomb-exploded by an intense
ultrashort probe-pulse, polarized parallel to the pump pulse, and
with a duration of typically some tens of femtoseconds. The
molecular fragments recoil approximately back-to-back and their
angular distribution can be found from the imaged fragments. Because
of the cylindrical symmetry, the true angular distribution can be
found even though the fragments are pulled by an electric field
towards a detection plane containing the laser polarizations. It is
the purpose of this paper to present a tomographic method to
reconstruct the rotational quantum state of the system from the
measurement data in such experiments.

We arrange the paper as follows: In section \ref{sec:qstintro} we
give a brief introduction to quantum state tomography. In section
\ref{sec:teori} we present the tomographic method for the rigid
linear rotor. In section \ref{sec:centrifug} we show how to extend
the tomographic method to molecules where centrifugal distortion
must be taken into account. In section \ref{sec:symtop} we extend
the tomographic method to the case of symmetric top molecules. In
section \ref{sec:diskussion} we discuss the possibilities of
recording the experimental data necessary to perform a
reconstruction and we conclude the paper.
%, corresponding to the operator
%$\int_0^{2\pi}{d\theta}|\theta,\phi\rangle_t
%{}_t\hspace{-0.03cm}\langle\theta,\phi|$.

\section{Quantum state tomography \label{sec:qstintro}}
Mathematically, tomography is a technique, closely related to
Fourier transformation, by which one can find a function from
knowing its projection along all rotated lines. It found its first
application in physiology in the 1970s, and tomographic techniques
are now used extensively in hospital scanners, where images of
internal tissue can be found from data recorded using X-rays or NMR.
Apart from physiology, tomography is currently also being used in
many other fields of research, including quantum physics and
chemistry. Here the aim is to use measurements to find the quantum
state of a system. The measurements are often taken to be spatial
probability distributions at different times, which for harmonic
oscillator systems makes the method very similar to the scanning
methods from physiology \cite{invradon}, \cite{invradon2}. Quantum
state reconstruction has also been considered for many other systems
including particles in traps (neutral atoms \cite{Buzekatom} and
ions \cite{Wineland}), general one-dimensional systems
\cite{LeonhardtRaymerprl}, dissociating molecules \cite{Juhl} and
the angular state of an electron in a Hydrogen atom with $n = 3$
\cite{nlig3angmomtom}.

\section{\label{sec:teori}Reconstruction method for linear molecules}
In this section we will present a reconstruction method by which one
can find the rotational state of a linear rigid rotor. As usually in
quantum tomography, we imagine that we are able to perform
measurements of spatial distributions at different points of time
and that we know the Hamiltonian governing the time evolution. The
goal shall be to find the unique quantum state corresponding to
these measurements.

\subsection{Free rotation}
Before discussing the description of the quantum state, we shall
first digress to account for what we know about the Hamiltonian and
its eigenstates. As discussed in the introduction, we are varying
the time elapsed between the rotational state is created by a pump
pulse and it is probed. During this time the molecule is in a
field-free environment. Consequently, we shall be interested in the
Hamiltonian for the free rotor.

The Hamiltonian governing the time evolution for the linear rotor
with moment of inertia $\mathcal{I}$ is:
\begin{eqnarray}
\hat H = \frac{\hat{\bm J}^2}{2\mathcal{I}}, \nonumber
\end{eqnarray}
where $\hat{\bm{J}}$ is the angular momentum operator. The energy
eigenstates of this Hamiltonian are $\{|J,m\rangle\}$ where the
angular momentum quantum number $J \in \mathbb{N}_0$ and the
projection quantum number $m \in \{-J,-J+1,\ldots,J\}$. These states
have energy $E_J = \hbar\Omega J(J+1)$, where $\Omega =
\frac{\hbar}{2\mathcal{I}}$. Since we will attempt to reconstruct
the quantum state using angular distribution measurements only, we
shall need the spatial coordinate representation of the energy
eigenstates. As discussed further below, we shall restrict our study
to azimuthally symmetric distributions, and we shall hence be
interested in the distribution of the polar angle $\theta$. For
notational simplicity, we shall use the parameter $x = \cos(\theta)$
so the position representation of the eigenstates becomes:
\begin{eqnarray}\label{eq:posket}
{}_t\langle x|J,m\rangle &=& \mathcal{P}_J^m(x)\,e^{-i\Omega J(J+1)
t}.
\end{eqnarray}
There should be a factor of $e^{im\phi}$ in Eq.~(\ref{eq:posket}),
but we can safely ignore this since we shall restrict our treatment
to fixed values of $m$ \footnote{Any phase factor will disappear
when multiplied with the complex conjugate in
Eq.~(\ref{eq:prxtj1j2}). Another way to look at this is that since
we assume azimuthal symmetry of the angular distributions, we might
as well evaluate the position distribution at the azimuthal angle
$\phi = 0$.}. The $\mathcal{P}_J^m$ are the normalized associated
Legendre polynomials, normalized on $x \in [-1,1]$
\cite{arfkenweber}:
\begin{eqnarray}
\int_{-1}^{1}\hspace{-0.39cm}dx\,\mathcal{P}^m_{J_1}(x)\mathcal{P}^m_{J_2}(x)
&=& \delta_{J_1,J_2}.
\end{eqnarray}
The multiplicative factors relating these to the un-normalized
associated Legendre polynomials $P_J^m(x)$ are:
\begin{eqnarray}
\mathcal{P}_J^m(x) &=&
\sqrt{\frac{2J+1}{2}\,\frac{(J-m)!}{(J+m)!}}P_J^m(x). \nonumber
\end{eqnarray}
Eq.~(\ref{eq:posket}) fully accounts for the free time evolution of
the rotor. The angular position distribution of a single eigenstate
is constant in time, but due to the different time dependent phase
factors, a linear combination of different $|J,m\rangle$ states will
have time dependent interference terms. Hence, such a linear
combination will have a time dependent position distribution, also
in the field free case.

After this brief account of the dynamics of the system, we proceed
to define what we mean by the quantum state we are trying to
reconstruct.

In the present treatment, the state will be characterized by its
density operator $\hat \rho$, which allows a description of
statistically mixed states. Specifically, we shall find all the
matrix elements of $\hat \rho$ in the $|J,m\rangle$-basis at $t = 0$
restricted to single $m$'s. This is the same as finding the diagonal
blocks of constant $m$ in the density matrix, i.e. $\rho(J,m,J',m) =
\rho_m(J,J')$. For a general state this is only a partial
characterization, but if it is somehow known that the state has no
correlations between different $m$ values, knowledge of all
$\rho_m(J,J')$ amounts to a full characterization of the state.
%The only
%assumption we make on the state is that there are no correlations
%between the different projection quantum numbers $m$, meaning that
%the density matrix is block diagonal in $m$.
This is the case in the experiment described in the introduction and
the procedure below will completely reconstruct the quantum state
for such systems. For the angular distribution, this diagonality in
$m$ implies cylindrical symmetry for all time around a laboratory
fixed axis, which we choose to be a spherical polar axis.
%As the azimuthal angular distribution is a constant, it is only
%necessary to measure the polar angular distribution at different
%points of time.

\subsection{Measurements and observables}
Having described the physical system, we proceed to consider the
measurements performed; namely the (angular) position distribution
measurements at the time $t$:
\begin{eqnarray}\label{eq:prxtj1j2}
\textnormal{Pr}(x,t) &=& {}_t\langle x|\hat\rho_m|x\rangle_t\nonumber \\
        &\hspace{-2cm}=&\hspace{-1.3cm}\sum_{J_1 = |m|}^{\infty}\sum_{J_2 = |m|}^{\infty}{}_t\hspace{-0.02cm}\langle
        x|J_1,m\rangle\langle J_1,m|\hat\rho_m|J_2,m\rangle\langle
        J_2,m|x\rangle_t \nonumber\\
        &\hspace{-2cm}=&\hspace{-1.3cm}\sum_{J_1 = |m|}^{\infty}\sum_{J_2 =
|m|}^{\infty}\rho_m\hspace{-0.09cm}\left(J_1,J_2\right)\mathcal{P}^m_{J_1}(x)\mathcal{P}^m_{J_2}(x) \times \nonumber\\
&&\hspace{-1.3cm}\qquad\qquad\quad\,\,\,
e^{-i\Omega\left[J_1(J_1+1)-J_2(J_2+1)\right]t},
\end{eqnarray}
where $t = 0$ is an arbitrary, but fixed, point of time after
passage of the pump pulse. Our aim is to invert this equation to
find the density matrix in the eigenstate-representation
$\rho_m\hspace{-0.02cm}(J_1,J_2)$. It will prove convenient to write
the products of the two Legendre polynomials
$\mathcal{P}^m_{J_1}(x)\mathcal{P}^m_{J_2}(x)$ as a sum of single
Legendre polynomials. This is exactly what is done in the
decomposition of direct product bases for irreducible
representations of the rotation group:
\begin{eqnarray}\label{eq:pptilp}
\hspace{-0.6cm}\mathcal{P}^m_{J_1}(x)\mathcal{P}^m_{J_2}(x)
&\hspace{-0.14cm}=& \hspace{-0.5cm}\sum_{L =
|J_1-J_2|}^{J_1+J_2}\hspace{-0.4cm}\sqrt{2\pi}\,C(J_1,J_2,L|m,-m,0)\mathcal{P}^0_L(x),
\end{eqnarray}
where $C(J_1,J_2,L|m,-m,0)$ are Clebsch-Gordan coefficients
\cite{arfkenweber}. Next, we introduce the variables $J = J_1+J_2$
and $\Delta J = J_1-J_2$, and the notational simplification:
\begin{eqnarray}
C_{J,\Delta J,L}^m &=&
\sqrt{2\pi}\,C\hspace{-0.07cm}\left(\frac{J+\Delta
J}{2},\frac{J-\Delta J}{2},L\Big|m,-m,0\right). \nonumber
\end{eqnarray}
Writing the position distributions Eq.~(\ref{eq:prxtj1j2}) in the
new variables $J$ and $\Delta J$, and using Eq.~(\ref{eq:pptilp}) we
arrive at:
\begin{eqnarray}\label{eq:prxtjj}
\textnormal{Pr}(x,t) &=& \left(\sum_{\substack{ J =
|m|\\\textnormal{$J$ even}}}^{\infty}\,\sum_{\substack{\Delta J = - J\\
\textnormal{$\Delta J$ even}}}^{J} + \sum_{\substack{ J =
|m| \\ \textnormal{$J$ odd}}}^{\infty}\,\sum_{\substack{\Delta J = - J\\
\textnormal{$\Delta J$ odd}}}^{
J}\right) \times \nonumber\\
&&\rho_m\hspace{-0.09cm}\left(\frac{ J + \Delta J}{2},\frac{ J - \Delta J}{2}\right) \times \nonumber\\
&& e^{-i\Omega\Delta J\left( J + 1\right)t}\sum_{L = |\Delta J|}^{
J}\mathcal{P}^0_{L}(x)C^m_{J,\Delta J,L}.
\end{eqnarray}
For readability, we shall in the following abbreviate the summations
in the first line of Eq.~(\ref{eq:prxtjj}) as:
\begin{eqnarray}\label{eq:eeoo}
\sum_{J,\,\Delta J}^{ee,oo}.
\end{eqnarray}
We have now expanded the time dependent position distribution on
orthogonal polynomials $\mathcal{P}_\alpha^0(x)$ with coefficieents
that can be found by simple spatial integrals $\int_{-1}^1
\hspace{-0.02cm}dx \mathcal{P}_\alpha^m(x)\textnormal{Pr}(x,t)$.
These coefficients, in turn, are linear combinations of the density
matrix elements which can be found if the distribution is recorded
at different times, as we show below.

\subsection{Reconstruction formulas}
In this subsection we will present the reconstruction formulas. We
will first consider the off-diagonal elements of the density matrix
and deal with the diagonal elements further below.

\subsubsection{Reconstruction of off-diagonal elements \label{sec:betanotnull}}
We let $N_T T$ be the time interval in which position measurements
$\textnormal{Pr}(x,t)$ have been performed. Here $N_T \in
\mathbb{N}$ and the minimum measurement time $T = 2\pi
\mathcal{I}/\hbar$, equalling $\sqrt{2}$ times the semiclassical
rotational period of the $J = 1$ state or the rotational revival
period \cite{rotrevival}.

The time dependent factor $e^{-i \Omega \Delta J (J+1)t}$ in
Eq.~(\ref{eq:prxtjj}) suggests that a temporal Fourier transform
evaluated at frequency $\Omega\beta(\alpha +1)$, with $\alpha$ and
$\beta$ suitably chosen integers, will be part of the reconstruction
procedure. Following this strategy, we consider the integral:

\begin{eqnarray}\label{eq:Ibegynd}
I(\alpha,\beta) = \frac{1}{N_T T}\int_{0}^{N_T
T}\hspace{-0.39cm}dt\,e^{i\Omega\beta(\alpha+1)t}\int_{-1}^{1}\hspace{-0.39cm}dx\,\mathcal{P}^0_\alpha(x)\textnormal{Pr}(x,t)&&\nonumber\\
%= \left(\sum_{\substack{ J =
%|m|\\ \textnormal{or} |m|+1\\ \textnormal{$J$ even}}}^{\infty}\sum_{\substack{\Delta J = - J\\
%\textnormal{$\Delta J$ even}}}^{J} + \sum_{\substack{J =
%|m|+1\\ \textnormal{or} |m|\\ \textnormal{$J$ odd}}}^{\infty}\sum_{\substack{\Delta J = - J\\
%\textnormal{$\Delta J$ odd}}}^{
%J}\right)
= \sum_{J,\,\Delta J}^{ee,oo} \rho_m\hspace{-0.09cm}\left(\frac{ J +
\Delta J}{2},\frac{J-\Delta
J}{2}\right)\delta_{\beta(\alpha+1)-\Delta J(J+1)}
\times \nonumber\\
\sum_{L=|\Delta J|}^{J}C^m_{J,\Delta J,L}\delta_{L,\alpha},\qquad&&
\end{eqnarray}
where $\alpha \in \{|m|,|m+1|,\ldots\}$ and $\beta \in \{-\alpha,
-\alpha+2,\ldots,\alpha\}$, which makes both $\alpha$ and $\beta$
are either even or odd. For the time being, we shall choose $\beta
\neq 0$ and deal with the $\beta = 0$ case (i.e. diagonal matrix
elements) later.

Notice that the delta-function in the sum over $L$ ensures that
$\alpha$, $J$ and $\Delta J$ are all of the same parity. This is
also the parity of $\beta$, since we chose $\alpha$ and $\beta$ to
be of the same parity.
\begin{eqnarray}\label{eq:Ifortsaet}
I(\alpha,\beta) &=& \hspace{-0.45cm}\sum_{\substack{J = |m|\\
\textnormal{parity as
$\alpha$}}}^{\infty}\sum_{\substack{\Delta J = - J\\
\textnormal{parity as
$\alpha$}}}^{J}\hspace{-0.15cm}\rho_m\hspace{-0.09cm}\left(\frac{J +
\Delta
J}{2},\frac{J - \Delta J}{2}\right)\times \nonumber\\
&&\delta_{\beta(\alpha+1)-\Delta J(J+1)}C^m_{J,\Delta J,\alpha}
\sum_{L=|\Delta J|}^{J}\delta_{L,\alpha}.
\end{eqnarray}
Rather than giving a single term in the general case, the sums are
greatly simplified. From the $L$-sum we get $I(\alpha,\beta) = 0$
unless:
\begin{eqnarray}\label{eq:beting1}
\left|\Delta J\right| \leq & \alpha & \leq J.
\end{eqnarray}
Using this in conjunction with the other delta-function we find
$I(\alpha,\beta) = 0$ unless:
\begin{eqnarray}\label{eq:beting2}
|\Delta J| \leq |\beta| \leq \alpha, \quad \textnormal{$\beta$ and
$\Delta J$ of same sign}.
\end{eqnarray}
Furthermore, this delta-function demands:
\begin{eqnarray}\label{eq:stordelta}
\beta\left(\alpha + 1\right) &=& \Delta J\left(J+1\right).
\end{eqnarray}
As it turns out, for many choices of $\alpha$ and $\beta$ there is
only the straightforward solution to the equation
Eq.~(\ref{eq:stordelta}) under the conditions
Eqs.~(\ref{eq:beting1}) and (\ref{eq:beting2}), namely $\alpha = J$
and $\beta = \Delta J$. In these cases we easily find the density
matrix elements:
\begin{eqnarray}\label{eq:Isimpel}
\hspace{-0.55cm}I\left(\alpha,\beta\right) &=&
C^m_{\beta,\alpha,\alpha}\,
\rho_m\hspace{-0.04cm}\hspace{-0.07cm}\left(\frac{\alpha +
\beta}{2},\frac{\alpha -
\beta}{2}\right),\,\substack{\textnormal{unique solution}\\
\textnormal{to \ref{eq:beting1}, \ref{eq:beting2} and
\ref{eq:stordelta}.}}
\end{eqnarray}
For some choices of $\alpha$ and $\beta$ there will be more than one
term surviving from the sums in Eq.~(\ref{eq:Ifortsaet}). The
physical reason for this is that there is more than one pair of
energy eigenstates having a certain energy difference. One example
is the energy difference between the states $(J_1=3\rightarrow
J_2=0)$ and between the states $(J_1=6\rightarrow J_2=5)$. In our
variables this corresponds to $(J=3,\Delta J = 3)$ and $(J = 11,
\Delta J = 1)$.
%These degeneracies have to be found numerically by using
%Eq.~(\ref{eq:stordelta}) and the conditions Eqs.~(\ref{eq:beting1})
%and (\ref{eq:beting2}).
We proceed to show that all density matrix elements with $\beta \neq
0$ can be found regardless of this complication.

Imagine that we have found $I(\alpha,\beta)$ to contain the term
$C^m_{\beta,\alpha,\alpha}\,\rho_m\hspace{-0.08cm}\left(\frac{\alpha
+ \beta}{2},\frac{\alpha - \beta}{2}\right)$, but also several other
terms $C^m_{\beta'_{n_1},\alpha'_{n_1},\alpha}
\,\rho_m\hspace{-0.08cm}\left(\frac{\alpha'_{n_1} +
\beta'_{n_1}}{2},\frac{\alpha'_{n_1} - \beta'_{n_1}}{2}\right)$,
where $n_1 = 1,2,\ldots,N_1$. The strategy is to calculate all the
integrals $I(\alpha_{n_1}',\beta_{n_1}')$. These may also contain
several terms with $(J =\alpha''_{n_2}, \Delta J = \beta_{n_2}'')$,
where $n_2 = 1,2,\ldots,N_2$, but always fewer than in the former
integrals, i.e. $N_j < N_{j+1}$. The reason for this is the
condition $|\Delta J| \leq |\beta|$ from Eq.~(\ref{eq:beting2}):
Whereas the $\beta = \Delta J$ term will always appear, the rest of
the terms in the sum Eq.~(\ref{eq:Ifortsaet}) will have $\Delta J <
\beta$, whereby the procedure will terminate and we can find all the
matrix elements of $\hat \rho_m$ by back substitution. If one is
interested only in matrix elements up to some maximum $J = J_{max}$
it is only necessary to search for degeneracies in the energy
differences for all $(J,\beta)$ up to $J = J_{max}(J_{max}+1)/2$ for
even $J_{max}$ and $J_{max}(J_{max+1})$ for odd $J_{max}$.

To clarify this procedure a little, let us consider an example: We
shall choose $m = 0$ and try to find the matrix element
$\rho_0\hspace{-0.04cm}(5,0)$. In this case $J = \Delta J = 5$, and
we find $I(5,5) =
C^m_{5,5,5}\rho_m\hspace{-0.04cm}(5,0)+C^m_{9,3,5}\rho_m\hspace{-0.04cm}(6,3)+C^m_{29,1,5}\rho_m\hspace{-0.04cm}(15,14)$.
To find the desired matrix element we shall need the two extra
equations $I(9,3) =
C^m_{9,3,3}\rho_m\hspace{-0.04cm}(6,3)+C^m_{29,1,3}\rho_m\hspace{-0.04cm}(15,14)$
and $I(29,1) = C^m_{29,1,1}\rho_m\hspace{-0.04cm}(15,14)$. These
three resulting equations can now be solved by back substitution to
yield $\rho_m\hspace{-0.04cm}(5,0)$, $\rho_m\hspace{-0.04cm}(6,3)$
and $\rho_m\hspace{-0.04cm}(15,14)$.

\subsubsection{Reconstructing the diagonal\label{sec:betanull}}
We proceed to show how one can go about finding the diagonal of
$\hat \rho_m$ in the energy-representation. To do this we shall use
Eq.~(\ref{eq:Ifortsaet}) with $\beta = 0$, by which $\alpha$ is
even:
\begin{eqnarray}\label{eq:Imaerke}
I(\alpha,0) &=& \frac{1}{N_T T}\int_{0}^{N_T
T}\hspace{-0.63cm}dt\,\int_{-1}^{1}\hspace{-0.34cm}dx\,
\mathcal{P}^0_{\alpha}(x) \textnormal{Pr}(x,t) \nonumber \\
&=& \sum^{\infty}_{\substack{J = |m|\\
\textnormal{$J$ even}}}\rho_m\hspace{-0.09cm}\left(\frac{J}{2},\frac{J}{2}\right)C^m_{J,0,\alpha}\nonumber\\
&=&
\sum^{\infty}_{J_1=|m|}\rho_m\hspace{-0.09cm}\left(J_1,J_1\right)C^m_{2J_1,0,\alpha},
\end{eqnarray}
where we have used $J_1 = J/2$. We notice that the coefficient
$C^m_{2J_1,0,\alpha}$ is zero unless $\alpha \leq 2J_1$. We arrange
the $I(\alpha,0)$ in a column vector $\bm I$ and the
$\rho_m\hspace{-0.04cm}(J_1,J_1)$ in a column vector $\bm R$:
\begin{eqnarray}\label{eq:IligMrho}
\bm{I} &=&
\left(\begin{array}{c} I\left[2|m|,0\right]\\
I\left[2(|m|+1),0\right]\\
%I'(|m|+2)\\
\vdots
\end{array}\right),\nonumber \\
\bm R &=& \left( \begin{array}{c}
\rho_m(|m|,|m|)\\
\rho_m(|m|+1,|m|+1)\\
%\rho_m(|m|+2,|m|+2)\\
\vdots \end{array}\right). \nonumber
\end{eqnarray}
Eq.~(\ref{eq:Imaerke}) now reads:
\begin{eqnarray}
\bm I &=& \mathscr{M}\cdot{\bm R}, \qquad
\textnormal{where}\\
\mathscr{M} &=& \underbrace{\left(\begin{array}{c
c c} C^m_{2|m|,0,2|m|} &
C^m_{2(|m|+1),0,2|m|} %& C^m_{2(|m|+2),0,2|m|}
& \ldots\\
0& C^m_{2(|m|+1),0,2(|m|+1)} & %C^m_{2(|m+2|),0,2(|m|+1)}&
 \ldots \\
%0&0& C^m_{2(|m|+2),0,2(|m|+2)}&\ldots \\
\vdots&\vdots & %\vdots &
\ddots
\end{array}\right)}_{upper-triangular} \nonumber
%&=& \mathscr{M}\cdot{\textnormal{diag}({\rho_m})}.
\end{eqnarray}
To find the desired elements $\rho_m\hspace{-0.04cm}(J_1,J_1)$ we
shall need to invert the matrix $\mathscr{M}$, which can be done
since the diagonal elements $C^m_{\alpha,0,\alpha} =
\sqrt{2\pi}C(\alpha/2,\alpha/2,\alpha|m,-m,0)$ are all non-zero. In
practical applications, one may truncate the linear system at some
maximum $J_1$ corresponding to a maximum energy.

It may be noted that since the linear system (\ref{eq:IligMrho})
always has the same coefficient matrix, we could just has well have
used a function $\mathcal{F}_{J_1}$ that selects a certain $J_1$ in
(\ref{eq:Imaerke}), so that:
\begin{eqnarray}
\rho_m(J_1,J_1)&=& \frac{1}{N_T T}\int_{0}^{N_T
T}\hspace{-0.63cm}dt\,\int_{-1}^{1}\hspace{-0.34cm}dx\,
\mathcal{F}_{J_1}(x) \textnormal{Pr}(x,t). \nonumber
\end{eqnarray}
The $\mathcal{F}_{J_1}(x)$ can be precalculated and are given by:
\begin{eqnarray}
\mathcal{F}_{J_1}(x) &=& \sum_{J = J_1}^\infty f_{J_1,J}
\mathcal{P}_{2J}(x), \nonumber
\end{eqnarray}
where the expansion coefficients $f_{J_1,J}$ are easily found by
Cramer's rule and the fact that $\mathscr{M}$ is upper-triangular:
\begin{eqnarray}
f_{J_1,J} =
(-1)^{J_1+J}\frac{\left|\mathscr{M}^{\textnormal{sub}}_{J_1,J}\right|}{\prod_{\gamma
= J_1}^{J}\mathscr{M}(\gamma,\gamma)},
\end{eqnarray}
and $\left|\mathscr{M}^{\textnormal{sub}}_{J_1,J}\right|$ is the
determinant of the sub-matrix:
\begin{eqnarray}
\mathscr{M}^{\textnormal{sub}}_{J_1,J} &=&
\mathscr{M}\left(J_1+1:J_1+1+J, J_1+2:J_1+2+J\right). \nonumber
\end{eqnarray}
This method would be relevant if only a few particular elements of
the density matrix are desired. One may notice the similarity of
these functions to the pattern functions used with harmonic
oscillator systems \cite{patternfunc},\cite{pattern2}.

It may be noted that we could not have circumvented the back
substitution procedure in section \ref{sec:betanotnull} by choosing
other functions than the $\mathcal{P}^0_\alpha(x)$ in
Eq.~(\ref{eq:Ifortsaet}), since the solution of the resulting system
of equations depends on the measurement values through the
$I(\alpha_j,\beta_j)$'s.

\section{\label{sec:centrifug}Real molecules and centrifugal
distortion} In the above treatment we used the rigid rotor
Hamiltonian $\hat H = \frac{\hat{\bm{J}}^2}{2\mathcal{I}}$, whereas
terms of higher order in $\hat{\bm J}^2$ will be important in real
molecules in highly excited rotational states. With the lowest order
centrifugal distortion the Hamiltonian is instead:
\begin{eqnarray}
\hat H' &=& \frac{\hat{\bm{J}}^2}{2\mathcal{I}} -
\frac{D\hat{\bm{J}}^4}{\hbar^3}. \nonumber
\end{eqnarray}
The states $|J,m\rangle$ are still the energy eigenstates, but here
with energy $E_J = \hbar \Omega J(J+1) - \hbar D J^2(J+1)^2$. The
constant $D \propto \frac{1}{\mathcal{I}^3 \omega^2}$, where
$\omega$ is the harmonic frequency of the molecular bond. The
reconstruction strategy is similar to the one in
Eq.~(\ref{eq:Ibegynd}):
\begin{eqnarray}\label{eq:Imedcent}
I_{CD}(\alpha,\beta) &=& \lim_{T \rightarrow
\infty}\frac{1}{T}\int_{0}^{T}\hspace{-0.39cm}dt\,e^{i\beta(\alpha+1)\left[\Omega-D\beta(\alpha+1)\right]t}\times\nonumber\\
&&\quad\qquad\int_{-1}^{1}\hspace{-0.39cm}dx\,\mathcal{P}^0_\alpha(x)\textnormal{Pr}(x,t).
\end{eqnarray}
In this case, the conditions Eqs.~(\ref{eq:beting1}) and
(\ref{eq:beting2}) still apply, but Eq.~(\ref{eq:stordelta}) is
replaced by:
\begin{eqnarray}\label{eq:centstordelta}
\beta(\alpha+1) - \frac{D}{\Omega}\beta^2(\alpha+1)^2 = \quad\nonumber\\
\Delta J(J+1) - \frac{D}{\Omega}{\Delta J}^2\left(J+1\right)^2.
\end{eqnarray}
Depending on the value of $D/\Omega$, there may be more than one set
of $(\Delta J,J)$ that satisfies the conditions
Eqs.~(\ref{eq:beting1})-(\ref{eq:beting2}) and
Eq.~(\ref{eq:centstordelta}), but it will usually be much fewer than
in the rigid rotor case. Consequently, one can often avoid the back
substitution procedure above and use $I_{CD}(\alpha,\beta) =
C^m_{\beta,\alpha,\alpha}\,
\rho_m\hspace{-0.04cm}\hspace{-0.07cm}\left(\frac{\alpha +
\beta}{2},\frac{\alpha - \beta}{2}\right)$.

\section{Symmetric top states\label{sec:symtop}}
Having treated the linear rotor case, we will now straightforwardly
generalize the results from section \ref{sec:teori} to symmetric top
molecules. This class contains many more molecules than the linear
class.

The eigenvectors of the Hamiltonian are in this case the set
$\{|Jkm\rangle\}$ with associated energy $E = \hbar\Omega_1 J(J+1) -
\hbar \Omega_2 k^2$. Here $k$ is the quantum number for the
projection of the angular momentum on the symmetry axis of the
molecule, where $m$ was associated with a projection on a
space-fixed axis. Like $m$, $k$ can assume the values
$-J,-J+1,\ldots,J$. For a linear molecule $k = 0$, and the treatment
in section \ref{sec:teori} can indeed be seen as a special case of
the symmetric top. In accordance with the above treatment we shall
assume that we can measure the polar angular distribution,
parametrized by $x = \cos(\theta)$, for a certain value of $k$ and
$m$. The angular position representations of the eigenstates are:
\begin{eqnarray}
{}_t\hspace{-0.02cm}\langle x|Jkm\rangle &=&
\sqrt{\frac{2J+1}{2}}d^J_{km}(x)e^{-i(\Omega_1 J(J+1) - \Omega_2
k^2)}, \nonumber
\end{eqnarray}
where the $d^{J}_{km}(x)$ are the usual rotation matrix elements,
and the eigenstates are normalized on $[-1,1]$. The quantities
$\Omega_1 = 1/2\mathcal{I}_x$ and $\Omega_2 =
\hbar/(2\mathcal{I}_z-2\mathcal{I}_x)$. Here we have chosen the $z$
axis as the molecular symmetry axis, and used that the moments of
inertia $\mathcal{I}_x = \mathcal{I}_y$. As in
Eq.~(\ref{eq:posket}), we can safely ignore the dependence of the
two other Euler angles $\phi$ and $\chi$, since we are working in
the subspace of fixed $k$ and $m$. We can now readily generalize the
treatment in section \ref{sec:teori} by exchanging the functions
$\mathcal{P}_J^m(x)$ with $\sqrt{(2J+1)/2}\,d^J_{km}(x)$ and the
lower summation indices $|m|$ with $M_{km} =
\textnormal{max}(|m|,|k|)$. In particular, Eq.~(\ref{eq:prxtj1j2})
generalizes to:
\begin{eqnarray}\label{eq:symprxtj1j2}
\textnormal{Pr}(x,t) &=& {}_t\langle x|\hat\rho_{k,m}|x\rangle_t\nonumber \\
        &\hspace{-2cm}=&\hspace{-1.3cm}\sum_{\substack{J_1 = \\M_{km}}}^{\infty}\sum_{\substack{J_2 =\\M_{km}}}^{\infty}
        \rho_m\hspace{-0.09cm}\left(J_1,J_2\right)d^{J_1}_{km}(x)d^{J_2}_{km}(x) \times \nonumber\\
&&\hspace{-1.3cm}
\sqrt{\frac{(2J_1+1)(2J_2+1)}{4}}e^{-i\Omega\left[J_1(J_1+1)-J_2(J_2+1)\right]t}.
\end{eqnarray}
Products of the functions $d^{J}_{km}$(x) can again be decomposed
into linear combinations of single Legendre polynomials, again
introducing $J = (J_1+J_2)/2$ and $\Delta J = (J_1-J_2)/2$
\cite{tinkham}, \cite{Sakurai-san}:
\begin{eqnarray}
\sqrt{\frac{(2J_1+1)(2J_2+1)}{4}} d^{J_1}_{km}(x)d^{J_2}_{km}(x) &=& \nonumber \\
%\sum_{L = |J_1-J_2|}^{J_1+J_2}
%\sqrt{\frac{2}{2L+1}}C(J_1,J_2,L|m,-m,0) \times&&\nonumber \\
%C(J_1,J_2,L;k,-k,0)\mathcal{P}_L^0(x) = &&\nonumber\\
\sum_{L = |J_1-J_2|}^{J_1+J_2}C^{km}_{J,\Delta
J,L}\mathcal{P}_L^0(x),
\end{eqnarray}
where
\begin{eqnarray}\label{eq:Ckm}
C^{km}_{J,\Delta J,L} &=& \sqrt{\frac{2}{2L+1}}C(J_1,J_2,L|m,-m,0) \times\nonumber \\
&&\,\qquad\qquad C(J_1,J_2,L;k,-k,0).
\end{eqnarray}
Proceeding as in section \ref{sec:teori}, we calculate the integrals
$I(\alpha,\beta)$ and arrive at the equivalent of
Eq.~(\ref{eq:Ifortsaet}):
\begin{eqnarray}\label{eq:Isym}
I(\alpha,\beta) &=& \hspace{-0.45cm}\sum_{\substack{J = M_{km} \\
\textnormal{parity as $\alpha$}}}^{\infty}\,\sum_{\substack{\Delta J= -J\\
\textnormal{parity as
$\alpha$}}}^{J}\hspace{-0.15cm}\rho_{k,m}\hspace{-0.09cm}\left(\frac{J
+ \Delta
J}{2},\frac{J - \Delta J}{2}\right)\times \nonumber\\
&&\delta_{\beta(\alpha+1)-\Delta J(J+1)}C^{km}_{J,\Delta J,\alpha}
\sum_{L=|\Delta J|}^{J}\delta_{L,\alpha}.
\end{eqnarray}
The reconstruction of the matrix elements $\rho_{k,m}(J_1,J_2)$ can
therefore be done like in sections \ref{sec:betanotnull} and
\ref{sec:betanull}, where one uses the $C^{km}_{J,\Delta J,L}$
defined in Eq.~(\ref{eq:Ckm}) instead of $C^{m}_{J,\Delta J,L}$.

\section{Discussion \label{sec:diskussion}}
As the reader may have noticed, the method above effectively deals
with a semi-continuous one-dimensional reconstruction in the space
of $x$ and $J$, and not with the reconstruction of a general state.
The reason for this limitation is that the inversion formulas we are
using are integral transformations that preserve dimensionality,
like e.g. the Fourier transformation in $t$ in
Eq.~(\ref{eq:Ifortsaet}). If the dimensionality of the measurements
are not at least as large as the dimensionality of the quantum
state, such an approach nearly always fails \cite{uspiselig}. For
example, in one dimension the density matrix is a two-dimensional
object $\langle x|\hat \rho_{k,m}|x'\rangle$ which has the same
dimensionality as the measurement-space $\textnormal{Pr}(x,t)$. In
contrast, the full quantum state of e.g. the linear rotor is a
four-dimensional object $\langle x,\phi|\hat \rho|x',\phi'\rangle$,
while the angular measurements $\textnormal{Pr}(x,\phi,t)$ would
only be three-dimensional. If one desired to reconstruct a full
general quantum state, one would either have to be able to perform
more advanced measurements or to vary the Hamiltonian as suggested
in \cite{raymervarham}. Because of the difficulty of precisely
knowing Hamiltonians arising from e.g. additional laser pulses, we
have instead treated the one-dimensional system using only routinely
performed measurements.

In spite of this limitation, the situation where the one-dimensional
description is a complete characterization of the quantum system is
experimentally common. Indeed, for the situation described in the
introduction, there are no correlations between the different $k$
and $m$ quantum numbers. The question is of course whether one can
perform measurements of polar angular distributions for certain $k$
and $m$ quantum numbers.

A conceptually straightforward solution would be to use very
low-temperature samples, where only $k = m = 0$ is populated before
the interaction with the pump pulse. Note that this would not imply
that the created rotational state would be a pure state due to
inhomogeneities in the pump pulse focus, e.g. the pump intensity
generally varies over the focus or shot-to-shot variation of laser
intensity. Though conceptually simple, this situation is unpractical
due to the smallness of the rotational energies demanding very low
temperatures.

A more practical approach would be to select a certain value of $k$
and $m$ before interaction with the pump pulse. For polar symmetric
top molecules this can be done for certain values of $k$ and $m$ by
using the linear Stark effect through hexapole focusing techniques
\cite{henriksref10}, \cite{hexia}. The linear Stark shift for
symmetric tops is ${\Delta E^{(1)} = -\mu \varepsilon k m/J(J+1)}$,
where $\mu$ is the dipole moment and $\varepsilon$ is the magnitude
of the electric field. Since one starts out with a thermal sample,
there are no correlations in either $J$, $k$ or $m$ in the state
before interactions with the pump. Therefore, one does not discard
any information on the state by picking out all sets of values of
$J$, $k$ and $m$ one by one. If the molecular beam was sufficiently
cold, one could select components of the beam with certain $k$ and
$m$ values, as is required by the method above. One would then also
select a certain value of $J$, which is not required, but on the
other hand poses no problem. Furthermore, the apparent ambiguity
that a change of sign of both $k$ and $m$ gives the same Stark shift
is of no consequence, since the state created by the pump has
identical matrix elements $\rho_{k,m}(J,J')$ and
$\rho_{-k,-m}(J,J')$ due to the invariance of the initial state
 and of the pump-pulse interaction Hamiltonians under the interchange
$(k,m) \rightarrow (-k,-m)$.

Most linear molecules and symmetric tops with $k = 0$ do not exhibit
any linear Stark shift \cite{Flemmingbog}, but also here hexapole
focusing techniques can be used to select specific values of $J$ and
$m$ through the second order Stark shift. Here, both molecules with
the quantum numbers $(J,m)$ and $(J,-m)$ are selected, but this is
of no consequence since $\rho_m(J,J') = \rho_{-m}(J,J')$ in accord
with the previous paragraph.

In conclusion, we have presented a method to reconstruct the blocks
$\rho_{k,m}(J,J')$ of the rotational density matrix for linear and
symmetric top molecules from angular distributions. This amounts to
a complete characterization of the quantum state in a common
experimental setup. Finally, we have suggested how the required
measurements could be performed with existing techniques.
% Rather than performing measurements on a single quantum system, we
%shall imagine having a vast ensemble of identically prepared and
%uncorrelated systems. On each of these we perform only one
%measurement, after which this ensemble member is discarded.
%Experimentally, this ensemble idea merely means that we must have a
%sufficiently large body of measurements to have statistically
%significant data, and that no two incompatible measurements are
%performed on any one partial system.
%\begin{eqnarray}
%\mathcal{P}_{J_1}^0(x)\mathcal{P}_{J_2}^0(x) &=& \sum^{J_1+J_2}_{k =
%|J_1-J_2|}b(J_1,J_2,k)\mathcal{P}_k^0(x)\nonumber\\
%b(J_1,J_2,k)
%C(J_1,J_1,k|0,0,0)&=&
%\delta_{0,\left[\frac{1}{2}(J_1+J_2+k)\right]\textnormal{mod}
%1}\sqrt{\frac{(2k+1)}{\pi(2J_1+1)(2J_2+1)}}\times\nonumber\\
%&&\hspace{-1.5cm}\frac{(2k+1)(k+J_1-J_2-1)!!(k-J_1+J_2-1)!!}{(k+J_1-J_2)!!(k-J_1+J_2)!!} \times \nonumber \\
%&&\hspace{-1.5cm}\frac{(-k+J_1+J_2-1)!!(k+J_1+J_2)!!}{(-k+J_1+J_2)!!(k+J_1+J_2+1)!!}
%\end{eqnarray}

\begin{acknowledgements}
We wish to thank associate professor Flemming Hegelund and associate
professor Henrik Stapelfeldt for enlightening discussions and Simon
S. Viftrup technical advice and for providing experimental data for
the figures.
\end{acknowledgements}

\end{document}